\documentclass[12pt,preprint]{aastex}
\begin{document}
\newcommand{\calu}{{\cal U}}
\newcommand{\calq}{{\cal Q}}
\newcommand{\bx}{{\rm \bf x}}
\newcommand{\bk}{{\bar{\kappa}}}
\title{Gravitational Lensing by Dark Matter Halos with Non-universal Density Profiles}
\author{Tong-Jie Zhang}
\affil{Department of Astronomy, Beijing Normal University, Beijing
100875, P.R.China; tjzhang@bnu.edu.cn}

\begin{abstract}
The statistics of gravitational lensing can provide us with a very
powerful probe of the mass distribution of matter in the universe.
By comparing predicted strong lensing probabilities with
observations, we can test the mass distribution of dark matter
halos, in particular, the inner density slope. In this letter,
unlike previous work that directly models the density profiles of
dark matter halos semi-analytically, we generalize the density
profiles of dark matter halos from high-resolution N-body
simulations by means of generalized Navarro-Frenk-White (GNFW)
models of three populations with slopes, $\alpha$, of about -1.5,
-1.3 and -1.1 for galaxies, groups and clusters, respectively.
This approach is an alternative and independent way to examine the
slopes of mass density profiles of halos. We present calculations
of lensing probabilities using these GNFW profiles for three
populations in various spatially flat cosmological models with a
cosmological constant $\Lambda$. It is shown that the compound
model of density profiles does not match well with the observed
lensing probabilities derived from the Jodrell-Bank VLA
Astrometric Survey data in combination with the Cosmic Lens
All-Sky Survey data. Together with the previous work on lensing
probability, our results suggest that a singular isothermal sphere
mass model of less than about $10^{13}h^{-1}M_{\sun}$ can predict
strong lensing probabilities that are consistent with observations
of small splitting angles.

\end{abstract}



\section{Introduction}

Mapping the mass distribution of matter in the universe has been a
major challenge for modern observational cosmology.  The only
direct procedure to weigh matter in the universe is measuring its
deflection of light by gravity. The statistics of gravitational
lensing can provide us with a very powerful probe of the mass
distribution of the Universe. By comparing predicted lensing
probabilities with observations, we can examine the mass
distributions of dark matter halos, in particular, their inner
density slopes. It is well known that the Jodrell-Bank VLA
Astrometric Survey (JVAS) and the Cosmic Lens All-Sky Survey
(CLASS)
\citep{2000IAUS..201E..47B,2003MNRAS.341....1M,2003MNRAS.341...13B}
have provided us with observations of strong lensing probabilities
for small image separations ranging from $0.3''$ to $3''$. Based
on the Cold Dark Matter (CDM) model, which has become the standard
theory of cosmic structure formation, the lensing probabilities
strongly depend on the density profiles of CDM halos. The lensing
model is usually described by a singular isothermal sphere (SIS),
the Navarro-Frenk-White (NFW) model
\citep{1996ApJ...462..563N,1997ApJ...490..493N}, or generalized
NFW (GNFW) density profiles of dark halos
\citep{1996MNRAS.278..488Z}. \cite{2002ApJ...566..652L} employed a
semi-analytical approach to analyze the gravitational lensing of
remote quasars by foreground dark halos and checked the
plausibility of various lensing models. They found that no model
can completely explain the current observations: the SIS models
predict too many lenses with large splitting angles, while the NFW
models predict too few small splitting angles. They therefore
further developed a two-population halo model for lensing: small
mass halos with a steep inner density slope and large mass halos
with a shallow inner density slope, concluding that a combination
of SIS and NFW halo models can reproduce the current observations
reasonably well.

Motivated by the sensitivity of the dependence of the image
separation distribution of lenses below $1^{''}$ on both the inner
mass profile of galactic halos and the faint end slope of the mass
and luminosity functions, \cite{2003ApJ...584L...1M} compared the
traditional approach that models lenses as SISs and the Schechter
luminosity function, with another method that invokes a certain
halo mass profile and the Press-Schechter mass function. She found
that dark matter halos cannot all be SISs, otherwise, the mass
function becomes steeper than the luminosity function and leads to
a relatively high lensing rate on smaller angular scales.

\cite{2003ApJ...595..603L} further proposed a lensing model based
on three populations of halos distinguished by halos mass:
Population 1 corresponds to normal galaxies of mass
$10^{10}h^{-1}M_{\sun}<M<10^{13}h^{-1}M_{\sun}$ with centers
dominated by baryonic matter and inner density profile slopes of
$\alpha=2$ (SISs); Population 2 corresponds to groups or clusters
of galaxies of mass $M>10^{13}h^{-1}M_{\sun}$ with centers of dark
matter and $\alpha=1.3$ (GNFW models); Population 3 corresponds to
dwarf galaxies or sub-galactic objects with mass
$M<10^{10}h^{-1}M_{\sun}$, whose centers lack baryons and are also
dominated by dark matter, and also described by GNFW models of
slope $\alpha=1.3$. Their results showed that both LCDM and OCDM
cosmological models are marginally consistent with current lensing
observations, while the SCDM model can be ruled out.

Besides these semi-analytical approaches, we can directly employ
the results of high-resolution simulations for the density
profiles of dark matter halos in the calculation of lensing
probabilities. \cite{2000ApJ...529L..69J} performed a series of
high-resolution N-body simulations to examine the density profiles
of dark matter halos. They found a clear systematic correlation
between halo mass and the slope of the profile at $1\%$ of the
virial radius, the slope of which is $\sim$-1.5, -1.3 and -1.1 for
profiles of galaxies, groups and clusters, respectively. They
concluded that dark matter density profiles, especially in their
inner regions, do not follow universal forms such as the NFW
model. Therefore, consideration of the dependence of the slopes of
density profiles on the masses of halos in calculating lensing
probabilities seems to be necessary and reasonable. In this
letter, we generalize \cite{2000ApJ...529L..69J}'s cosmological
simulation results for density profiles of dark matter halos using
the three-population GNFW models with slopes of $\alpha\propto$
-1.5,-1.3 and -1.1 for galaxies, groups and clusters,
respectively. This is an alternative and independent way to
examine the slopes of density profiles of CDM halos.

\section{General Formalism for Lensing}

\subsection{Lensing Equation for GNFW models}

\begin{figure}
\epsscale{0.8} \plotone{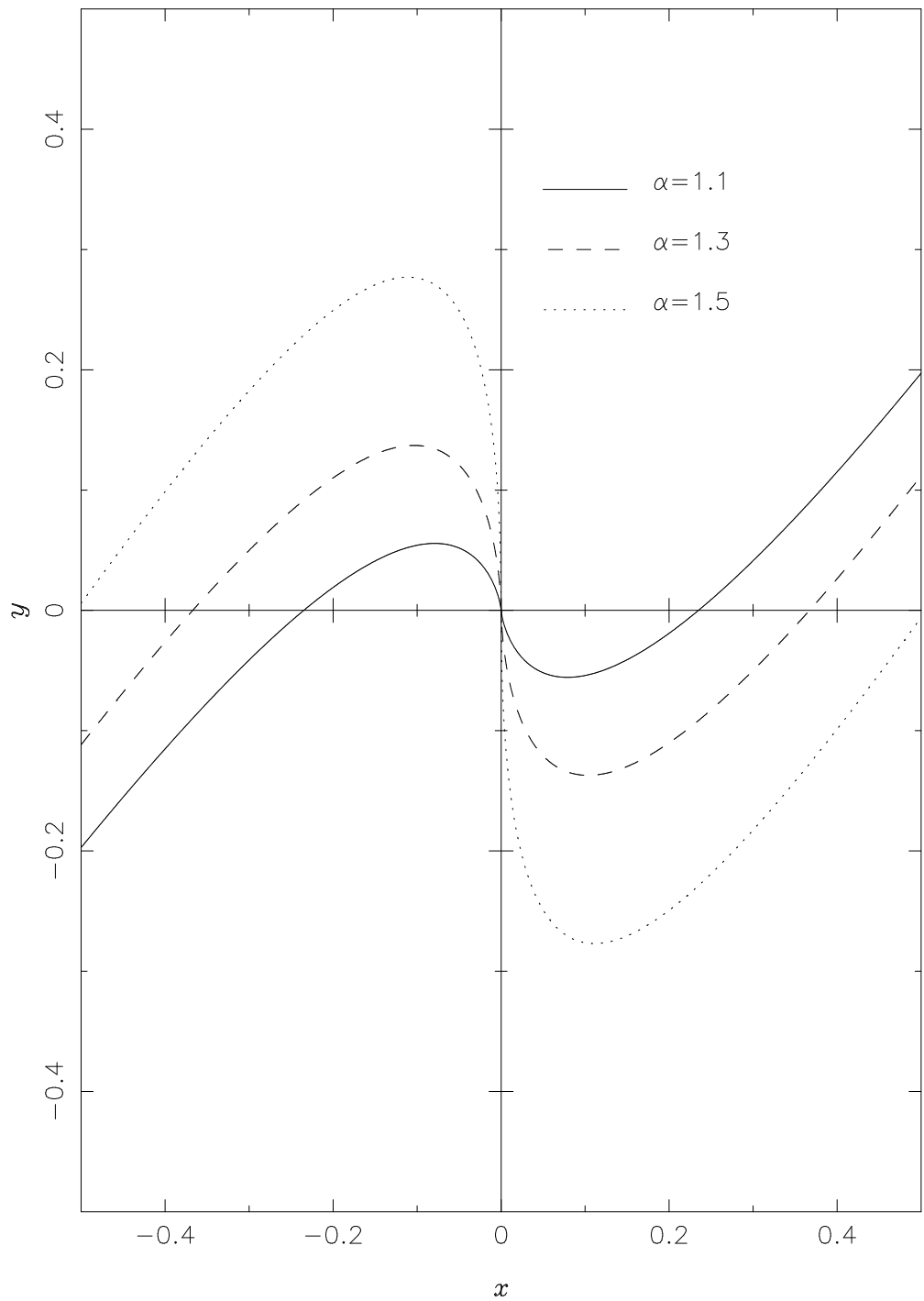} \caption{The lensing equation for
a GNFW model with $\alpha$=1.5, 1.3 and 1.1 respectively.}
\label{fig:le}
\end{figure}

The GNFW density profile can be expressed in the form
$\rho(r)={\rho_s r_s^3/r^{\alpha}(r+r_s)^{3-\alpha}}$
\citep{1996MNRAS.278..488Z} where $0<\alpha<3$. We generalize the
inner slope from the simulation of dark matter halos using the
GNFW model with a varying $\alpha$ which is dependent of the
masses of halos. We take $\alpha=1.5$ for $M<M_{c1}$, $\alpha=1.3$
for $M_{c1}<M<M_{c2}$ and $\alpha=1.1$ for $M>M_{c2}$,
respectively, where $M_{c1}\sim10^{13}M_{\sun}$ corresponds to the
cooling mass scale \citep{2000ApJ...532..679P,2001ApJ...559..531K}
and $M_{c2}\sim10^{14}M_{\sun}$. The mass of a dark halo within
$r_{200}$ can be defined as $M=4\pi\int^{r_{200}}_0\rho
r^2dr=4\pi\rho_\mathrm{s}r_\mathrm{s}^3f(c_1)$, and $r_{200}$ is
the radius of a sphere around a dark halo within which the average
mass density is $200$ times the critical mass density of the
universe. The function $f(c_1)=\int_0^{c_1} {x^2 dx/x^\alpha
(1+x)^{3-\alpha}}$ and $c_1=r_{200}/r_\mathrm{s}$ is the
concentration parameter
\begin{eqnarray}
    c_1(M_{15},z)=c_{\rm norm}
    \frac{2-\alpha}{1+z}[10M_{15}]^{-0.13},
    \label{c1}
\end{eqnarray}
where $z$ is halo redshift, $c_{\rm norm}=8$
\citep{2001MNRAS.321..559B} and $M_{15}=M/(10^{15}\mathrm{h}
^{-1}M_{\sun})$ is dimensionless halo mass. So $\rho_\mathrm{s}$
and $r_\mathrm{s}$ can be related to mass $M_{15}$ and redshift
$z$ by

\begin{equation}
\rho_\mathrm{s}=\rho_\mathrm{c_0}
E^2(z)\frac{200}{3}\frac{c_1^3}{f(c_1(M_{15},z))},\
r_\mathrm{s}=\frac{1.626}{c_1}\frac{M_{15}^{1/3}}
{E^{2/3}(z)}h^{-1}\mathrm{Mpc},
\end{equation}
where $\rho_{c_0}$ is the critical mass density of the universe
today and the expansion rate $E(z)=\sqrt{\Omega_\mathrm{m}
(1+z)^3+\Omega_{\Lambda}}$. The lensing equation for the GNFW
profile is given by $y=x-\mu_s g(x)/x$; $\vec{\xi}=\vec{x}r_s$ and
$\vec{\eta}=\vec{y}r_s d^A_S/d^A_L$ are the position vectors in
the lens plane and the source plane respectively, $g(x) \equiv
\int_0^x u du \int_0^{\infty} \left(u^2 +z^2\right)^{-\alpha/2}
\left[\left(u^2+z^2\right)^{1/2} +1\right]^{-3+\alpha} dz$, and
\begin{equation}
\mu_s\equiv{4\rho_s r_s\over \Sigma_{\rm
cr}}=0.002(\frac{\rho_s}{\rho_{c_0}})({r_s\over 1 h^{-1} {\rm
Mpc}})({d^A_R\over c/H_0}), \label{alps}
\end{equation}
where $\mu_s$ is a parameter on which the efficiency of producing
multiple images is strongly dependent. Here $\Sigma_{\rm
cr}={c^2\over 4\pi G}\,{d^A_S\over d^A_L d^A_{LS}}$ is the
critical surface mass density, and $d^A_R=d^A_L d^A_{LS}/d^A_S$.
$d^A_S$ and $d^A_L$ are the angular diameter distances from the
observer to the source and to the lens object respectively, while
$d^A_{LS}$ is the same quantity but from the lens to the source
object. The lensing equation for a GNFW density profile with
different $\alpha$ and setting $\mu_s=1$ is plotted in
Fig.\ref{fig:le}. The curves are symmetrical with respect to the
origin. Multiple images can be formed when $|y|\leq y_{\rm cr}$,
where $y_{\rm cr}$ is the maximum value of $y$ when $x<0$ or the
minimum value for $x>0$. Generally speaking, there exist three
images for $|y|< y_{\rm cr}$. We will just consider the outermost
two images stretched by the splitting angle $\Delta \theta$ when
more than two images are formed.

\subsection{Lensing Probability for GNFW models}

We can write the cross-section as $\sigma\left(M,z\right) \approx
\pi y_{\rm cr}^2 r_s^2\,\vartheta\left(\Delta\theta -
\Delta\theta_0\right)$ with $\Delta\theta>\Delta\theta_0$ in the
lens plane for multiple images produced by a GNFW lens at $z$.
$\vartheta$ is a step function and the splitting angle
$\Delta\theta$ is given by $\Delta\theta={r_s\Delta x/d^A_L}
\approx{2 x_0 r_s/d^A_L}$ where $x_0$ is the positive root of the
lensing equation $y(x)=0$. The lensing probability with image
separations larger than $\Delta\theta$ is given by
\cite{1992grle.book.....S}

\begin{equation}
P(>\Delta\theta)=\int^{z_{\mathrm{s}}}_0\frac{dD_{\mathrm{p}}(z)}
{dz}dz\int^{\infty}_0\bar{n}(M,z)\sigma(M,z)dM, \label{prob1}
\end{equation}
where $D_{\mathrm{p}}(z)=c/H_0\int^z_0 dz/(1+z)E(z)$ is the proper
distance from the observer to the lens at redshift $z$. The
physical number density $\bar{n}(M,z)$ of virialized dark halos of
masses between $M$ and $M+dM$ is expressed as
$\bar{n}(M,z)=n(M,z)(1+z)^3$, where $n(M,z)=\rho_0 f(M,z)/M$ is
the comoving number density \citep{1974ApJ...187..425P}. $\rho_0$
is the mean mass density of the universe today, and
$f(M,z)=-\sqrt{\frac{2}{\pi}}\frac{\delta_c(z)}{M\Delta}
\frac{d\ln\Delta}{d\ln M}\exp[-\frac{\delta_c^2(z)}{2\Delta^2}]$
where $\Delta^2(M)=\frac{1}{2\pi^2}\int^{\infty}_0P(k)
W^2(kr_{\mathrm{M}})k^2dk$ is the present variance of the
fluctuations within a sphere containing a mass $M$. The power
spectrum of CDM density fluctuations $P(k)=AkT^2(k)$ is given by
\cite{1999ApJ...511....5E}, where $A$ is the amplitude normalized
to $\sigma_8=\Delta (r_{\mathrm{M}}=8h^{-1}\mathrm{Mpc})$.

Since the redshift distribution of quasars in the JVAS/CLASS
survey is still poorly known, we adopt the mean value of
$z_s=1.27$ estimated by \cite{2000AJ....119.2629M}. In this
letter, we will use spatially flat $\Lambda$CDM models
characterized by the matter density parameter $\Omega_{\mathrm
m}$, vacuum energy density parameter $\Omega_{\Lambda}$ and Hubble
constant $h=0.75$, and will calculate lensing probabilities with
image separations greater than $\Delta\theta$ according to GNFW
halo profile models with varying $\alpha$. In the definition of
cross-section or lensing probability, we will just consider the
criterion $\Delta\theta$, and neglect another one $q_r$ (the
brightness ratio of the outermost two images) which is the ratio
of the corresponding absolute values of the magnifications
\citep{1992grle.book.....S}. In order to investigate the effect of
central black holes or bulges on lensing probability,
\cite{2003ApJ...587L..55C,2003A&A...397..415C} introduced $q_r$
into the calculation for lensing cross-section. Due to the
existence of central black holes or galactic bulges, $y_{cr}$
becomes extremely large when $|x|$ approaches zero. Thus $y_{cr}$
can be determined by the consideration of $q_r$ together with
$\Delta\theta$. However for GNFW halo models in the absence of
central black holes or galactic bulges, we can see in
Fig.\ref{fig:le} that the lensing equation curves are so smooth
that we do not need to define cross-section by $q_r$.

\begin{figure}
\epsscale{0.7} \plotone{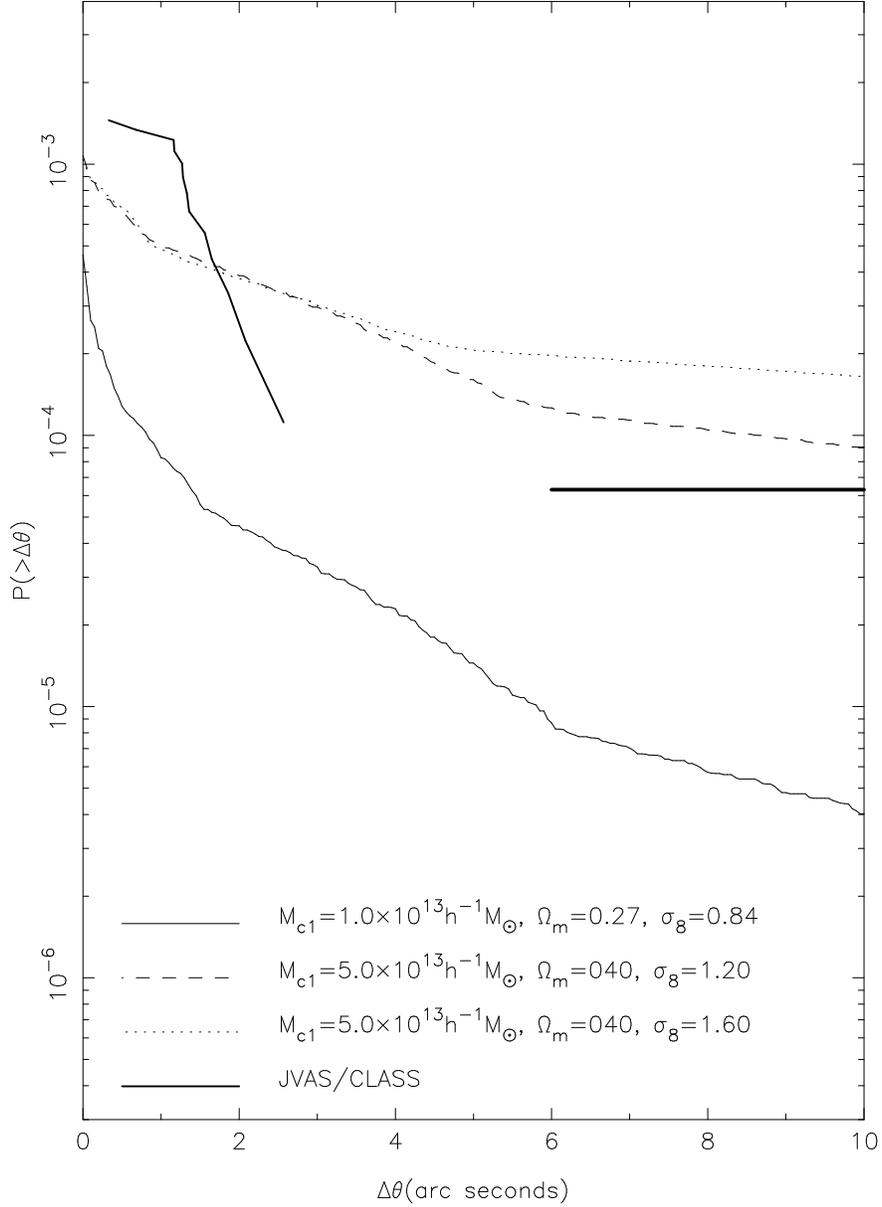} \caption{The lensing probabilities
for image separations greater than $\Delta\theta$ using the
compound GNFW dark halo profile models with varying $\alpha$. The
solid-line histogram represents observed lensing probabilities
from JVAS/CLASS \citep{2003ApJ...587L..55C,2002PhRvL..89o1301C}.
The thick solid horizontal line denotes the upper limit of the
null result for lenses with $6^{''}\leq\Delta\theta\leq 15^{''}$
from JVAS/CLASS. The other three curves are the predicted lensing
probabilities in $\Lambda$CDM cosmology.} \label{fig:lp}
\end{figure}

Our numerical results are shown in Fig.\ref{fig:lp} together with
the observational ones, which for $6^{''}\leq\Delta\theta\leq
15^{''}$ from JVAS/CLASS takes the form of an upper limit. Based
on the high-resolution simulation results
\citep{2000ApJ...529L..69J}, we adopt a compound GNFW model for
dark halos with three populations as described above. For the
cosmological models, we first choose the new result from the
Wilkinson Microwave Anisotropy Probe (WMAP): $\Omega_{\mathrm
m}=0.27$, $\sigma_8=0.84$
\citep{2003ApJS..148....1B,2003ApJS..148..175S} and
$M_{c1}=10^{13}M_{\sun}$, the numerical result of which is
depicted by a thin solid line in Fig.\ref{fig:lp}. It is clear
that the predicted lensing probabilities are much lower than the
observed values. Previous works
\citep{2003ApJ...587L..55C,2003A&A...397..415C} showed that
lensing probability is sensitive to $\Omega_{\mathrm m}$,
$\sigma_8$ and $M_{c1}$. More specifically, it increases with
$\Omega_{\mathrm m}$, $\sigma_8$ and $M_{c1}$. In order to improve
the prediction, we take $\Omega_{\mathrm m}$, $\sigma_8$ and
$M_{c1}$ to be the higher values of 0.4, 1.2 and
$5\times10^{13}M_{\sun}$ respectively. As shown by the dashed line
in Fig.\ref{fig:lp}, the derived lensing probabilities are still
inconsistent with the observations although they have increased
greatly compared to the former model. Thus we take the upper limit
of 1.6 for $\sigma_8$, as used in literature to date without
changing the other parameters. The results are shown with a doted
line in Fig.\ref{fig:lp}. There is almost no improvement in the
small angle separations compared with the case of $\sigma_8=1.2$,
apart from the increase of the predicted probabilities on large
angles. On the other hand, the lensing probabilities for both of
cases $\sigma_8=1.2$ and $\sigma_8=1.6$ lie above the observed
upper limit for large angle separations.

\section{Conclusions and Discussion}
We have generalized the density profiles from high-resolution
simulations of dark matter halos according to GNFW models with
varying $\alpha$, and developed a compound model incorporating
three populations of matter halos. We choose three typical
cosmological models: Model A, the new result of WMAP for model
parameters: $\Omega_{\mathrm m}=0.27$, $\sigma_8=0.84$ and
$M_{c1}=10^{13}M_{\sun}$; Model B, in which $\Omega_{\mathrm
m}=0.4$ is the upper limit adopted for the $\Lambda$CDM model,
$\sigma_8=1.2$ and $M_{c1}=5\times 10^{13}M_{\sun}$; and Model C,
which is the same as Model B except that $\sigma_8$ is taken to be
1.6, the maximum value used so far in the literature. We have
compared our predicted results for lensing probabilities with
JVAS/CLASS observations. Current observations for lensing
probabilities from JVAS/CLASS only cover lensing events on small
angle scales, though they also give the upper limit on separation
angles over the range 6$^{''}$ to 15$^{''}$. Our numerical results
as shown in Fig.\ref{fig:lp} show that the predicted lensing
probabilities of the halos of three populations in current
cosmological models do not agree with the JVAS/CLASS observations.
For Model A, the predicted probabilities of lensing events lie
below the upper limit observed on large angle scales, but is lower
than the observational results by a factor of about one order of
magnitude on small separation angles. Even when the upper limit
value of $\Lambda$CDM cosmology are used for $\Omega_{\mathrm m}$
and $\sigma_8$, such as in Models B and C, the predicted lensing
probability on small angle scales still do not match the
observations. In the calculation of cross section, we have not
considered the uncertainties such as the distribution of source
redshifts for quasars, the brightness ratio $q_r$, the scatter
effect of the concentration parameter $c_1$ and the variation in
$M_{c2}$. As discussed above, consideration of the brightness
ratio $q_r$ does not change our conclusion. As for the scatter in
the concentration parameter $c_1$,
\cite{2003ApJ...587L..55C,2003A&A...397..415C} have investigated
its effect on the prediction of lensing probabilities. By
averaging lensing probabilities with a log-normal distribution,
they found that the scatter in $c_1$ only increases the
probabilities on larger image separations but has little effect at
smaller separations. As for $M_{c2}$, any variation only
contributes to the predicted lensing probabilities on large angle
scales. In addition, for the sources in the JVAS/CLASS survey, the
redshift distribution is still poorly understood except for the
mean redshift $<z_s>$=1.27 as estimated by
\cite{2000AJ....119.2629M}. Thus, most of the previous work
\citep{2002ApJ...566..652L,2003ApJ...587L..55C} does not consider
the redshift distribution of the sources. The prediction of
\cite{1990MNRAS.247...19D} model and the CLASS lensing sub-sample
redshift measurements suggest that the redshift distribution for
CLASS unlensed sources can be modelled by a Gaussian distribution
with mean redshift $<z_s>$=1.27. However this can not affect the
conclusion of our letter. Even so, with the ever increasing number
of observed lensing events, we look forward to the day when the
redshift distribution derived from lensing observations can be
taken into account for other aspects of lensing study.

On the other hand, \cite{2001ApJ...554..903K} have presented a
convergence study of the density profiles of cold dark matter
halos simulated with varying mass and force resolutions. They
concluded that, on radii larger than the ``effective'' spatial
resolution, the density profiles of dark matter still do not
experience any systematical trends even with the further increase
of the number of particles or the force resolution. In addition,
they also explained the systematic correlation between inner
profile slope and halo mass found by \cite{2000ApJ...529L..69J},
and pointed out that it is a trend in dark matter halo
concentration. \cite{2003astro.ph.11361N} report recent results of
numerical simulations designed to study the inner slopes of
density profiles for CDM halos. Their results indicate that the
inner slopes of CDM halos are considerably shallower than the
asymptotic value of -1.5 proposed by \cite{1999ApJ...524L..19M}.
This point probably indirectly strengthens our argument that the
density profiles given by pure numerical simulation for CDM halos
are inconsistent with the observational statistics of strong
lensing. Therefore, in order to recover the observed strong
lensing probabilities from the inner density profiles of mass
halos, we can employ the SIS profile with a mass of less than
about $10^{13}h^{-1}M_{\sun}$. This is in agreement with previous
work such as \cite{2002ApJ...566..652L} and references therein.

This work was supported by the National Science Foundation of
China under grant No.10273003. I am very grateful to the anonymous
referee for many valuable comments that greatly improved the
paper. I would also like to thank Da-Ming Chen for useful
discussions, Ke-shih Young for much work in editing the
manuscript, Xiang-Ping Wu, Bo Qin, Ue-Li Pen and Peng-Jie Zhang
for their hospitality during my visits to the cosmology groups of
the National Astronomical Observatories of P.R.China and the
Canadian Institute for Theoretical Astrophysics(CITA), University
of Toronto.

\bibliography{ztjslensbib}
\bibliographystyle{apj}

\appendix

\end{document}